\documentclass[a4paper,11pt]{article}
\usepackage{pos}
\usepackage{booktabs}
\usepackage{wrapfig}
\usepackage{physics}
\usepackage{hyperref}
\usepackage{cleveref}
\usepackage{comment}
\usepackage[export]{adjustbox}
\usepackage{dsfont}
\usepackage{mathrsfs}

\newcommand{\gfi}{\gamma_5}
\newcommand{\gmu}{\gamma_\mu}
\newcommand{\amu}{a_\mu}
\newcommand{\HVP}{\mathrm{HVP}}
\newcommand{\mev}{\mathrm{MeV}}
\newcommand{\fm}{\mathrm{fm}}
\newcommand{\OS}{\mathsf{OS}}
\newcommand{\tm}{\mathsf{tm}}
\newcommand{\iso}{\mathsf{iso}}

\let\OLDthebibliography\thebibliography
\renewcommand\thebibliography[1]{
  \OLDthebibliography{#1}
  \setlength{\parskip}{1.4pt}
  \setlength{\itemsep}{0.6pt plus 0.5ex}
}

\begin{document}

\title{An update on the HVP contribution to $g_{\mu}{-}2$ \\ in isoQCD from ETMC}
\ShortTitle{An update on the HVP contribution to $g_{\mu}{-}2$ in isoQCD from ETMC}

\author*[a]{Antonio Evangelista}
\author[b]{Simone Bacchio}
\author[c, d]{Alessandro De Santis}
\author[e]{Roberto Frezzotti}
\author[f]{Giuseppe Gagliardi}
\author[g]{Marco Garofalo}
\author[e]{Lorenzo Maio}
\author[e]{Francesca Margari}
\author[b]{Ferenc Pittler}
\author[h]{Simone Romiti}

\affiliation[a]{Department of Physics, University of Cyprus, P.O. Box 20537, 1678 Nicosia, Cyprus}
\affiliation[b]{Computation-based Science and Technology Research Center, The Cyprus Institute, 20 Konstantinou Kavafi Street, 2121 Nicosia, Cyprus}
\affiliation[c]{Helmholtz-Institut Mainz, Johannes Gutenberg-Universit\"at Mainz, 55099 Mainz, Germany}
\affiliation[d]{GSI Helmholtz Centre for Heavy Ion Research, 64291 Darmstadt, Germany}

\affiliation[e]{Dipartimento di Fisica and INFN, Università di Roma “Tor Vergata", Via della Ricerca Scientifica 1, I-00133 Rome, Italy}
\affiliation[f]{Istituto Nazionale di Fisica Nucleare, Sezione di Roma Tre, Via della Vasca Navale 84, I-00146  Rome, Italy}
\affiliation[g]{HISKP (Theory), Rheinische Friedrich-Wilhelms-Universit\"at Bonn, Nussallee 14-16, 53115 Bonn, Germany}
\affiliation[h]{Institute for Theoretical Physics, Albert Einstein Center for Fundamental Physics, University of Bern, CH-3012 Bern, Switzerland}

\onbehalf{for the Extended Twisted Mass Collaboration (ETMC)}

\abstract{
We present an update on the determination of the leading-order hadronic vacuum polarisation contribution to the muon anomalous magnetic moment in isospin-symmetric QCD by the Extended Twisted Mass Collaboration. The calculation is based on five $N_f{=}2{+}1{+}1$ gauge ensembles generated with Wilson–clover twisted-mass quarks at maximal-twist and near-physical pion masses, spanning four lattice spacings and two volumes. For the dominant quark-connected contributions, we employ two distinct valence-quark regularisations and present results for both the isovector and isoscalar components.
}

\FullConference{The 42nd International Symposium on Lattice Field Theory (LATTICE2025)\\
2-8 November 2025\\
Tata Institute of Fundamental Research, Mumbai, India\\}
\maketitle

\setlength{\parskip}{7pt}
\setlength{\parindent}{0pt}

\section{Introduction}

The hadronic vacuum polarisation (HVP) contribution to the muon anomalous magnetic moment, $a_\mu^{\mathrm{HVP}}$, is the second-largest contribution after the purely QED term and represents one of the dominant sources of theoretical uncertainty. Dispersive evaluations have long exhibited a tension with the experimental measurements at Fermilab and Brookhaven~\cite{Aoyama:2020ynm,Aliberti:2025beg,Muong-2:2006rrc,Muong-2:2025xyk}, motivating first-principles determinations using lattice QCD. In recent years, several lattice collaborations have achieved subpercent precision~\cite{Borsanyi:2020mff,Boccaletti:2024guq,ExtendedTwistedMass:2022jpw,ExtendedTwistedMass:2024nyi,Djukanovic:2024cmq,MILC:2024ryz,FermilabLatticeHPQCD:2024ppc,RBC:2023pvn,RBC:2024fic}, finding results consistent with experiment and suggesting potential issues in the dispersive analyses.

In this contribution, we present the current status of the Extended Twisted Mass (ETM) Collaboration determination of $a_\mu^{\mathrm{HVP}}$ in isosymmetric QCD (isoQCD). Using gauge ensembles with $N_f = 2+1+1$ dynamical quarks generated with Wilson twisted-mass fermions at maximal twist, which ensures automatic $\mathcal{O}(a)$ improvement, we compute both isovector and isoscalar contributions. The analysis is performed on five ensembles spanning four lattice spacings and two physical volumes.

\section{The ETM Collaboration strategy}

In the time--momentum representation~\cite{Blum:2002ii,Bernecker:2011gh}, the leading-order HVP contribution reads
\begin{equation}
a_\mu^{\mathrm{HVP,\,LO}}
= 2\alpha^2 \int_0^\infty dt\, t^2\,K(t)\, C(t), \qquad
C(t) = -\frac{1}{3}\sum_{k=1}^{3}\int\dd[3]{\vb*{x}} \ev{\Hat{V}_{k}(\vb*{x}, t)\,\Hat{V}_{k}(\vb*{0}, 0)}{0}\,,
\label{eq:amu_def}
\end{equation}
where $\alpha$ is the QED coupling, $K(t)$ is a known kernel, and $C(t)$ is the zero-spatial-momentum Euclidean vector--vector two-point function. Neglecting the bottom-quark contribution, the electromagnetic current is
\begin{equation}
\Hat{V}_{\mu}(x) = \sum_{f} \Hat{V}_{\mu}^f(x), \qquad
\Hat{V}_{\mu}^f(x) = Z_V\, e_f\, \Bar{q}_f(x) \gmu q_f(x)\,,
\end{equation}
where $e_{f}$ is the electric charge of quark flavour $f=\{u,d,s,c\}$ in units of the positron charge.

We analyse data computed on five $N_f=2+1+1$ gauge ensembles generated by the ETM Collaboration with Wilson twisted-mass fermions at maximal twist~\cite{Frezzotti:2003ni, Frezzotti:2004wz, Alexandrou:2018egz, Finkenrath:2022eon}. The ensembles are tuned close to the isoQCD definition of the FLAG/Edinburgh consensus point~\cite{FlavourLatticeAveragingGroupFLAG:2024oxs},
\begin{equation}
f_\pi = 130.5~\mev, \qquad
m_\pi = 135~\mev, \qquad
m_K = 494.6~\mev, \qquad
m_{D_s} = 1967~\mev .
\end{equation}
The simulations cover lattice spacings in the range $a \simeq 0.079$--$0.050~\mathrm{fm}$ and spatial extents $L \simeq 5.1$--$7.6~\mathrm{fm}$ (see \cref{tab:iso_EDI_FLAG} for the parameters details).
The small residual mistuning from the isoQCD FLAG/Edinburgh point is quantified and controlled by analysing the derivative of the renormalisation observable with respect to both the twisted and critical masses.\footnote{
A detailed discussion of these mistunings will be presented in a forthcoming publication.
}

We work in a mixed-action setup in which the valence action (see Appendix~A of Ref.~\cite{ExtendedTwistedMass:2024nyi,ExtendedTwistedMass:2022jpw} for details) is
\begin{equation}
S_\OS(m_f,m_\mathrm{cr})
=
\sum_{f}\sum_x \bar{q}_f\left\{
\gmu \bar{\nabla}_\mu[U]
- i r_f \gfi \left(
W^\mathrm{cl}[U]+m_\mathrm{cr}
\right)
+ m_f
\right\}q_f \;,
\label{eq:OS_valence_quark_action}
\end{equation}
with $m_\ell=m_u=m_d$ and $r_f=\pm 1$. This setup allows us to define two distinct regularisations of the vector current that differ only by lattice artefacts,
\begin{equation}
\Hat{V}_{\mu}^{f, \tm}(x) = Z_V^{\tm}\, e_f\, \Bar{q}_f^{\pm}(x) \gmu q_f^{\mp}(x)\,, 
\qquad
\Hat{V}_{\mu}^{f, \OS}(x) = Z_V^{\OS}\, e_f\, \Bar{q}_f^{\pm}(x) \gmu q_f^{\pm}(x)\,,
\end{equation}
where the superscripts $\pm$ denote the sign of $r_f$ for flavour $f$, and $Z_V^{\mathsf{reg}}=Z_A$ in the $\tm$ case while $Z_V^{\mathsf{reg}}=Z_V$ in the $\OS$ case (see \cref{tab:iso_EDI_FLAG} for their values for each ensemble). Consequently, the connected Wick contraction of the vector--vector two-point correlator can be computed in two independent regularisations, whereas the disconnected contribution is evaluated in a single regularisation. This provides a useful handle to assess discretisation effects in the connected component as the continuum limit is approached.

Following the strategy proposed by the CLS/Mainz collaboration~\cite{Djukanovic:2024cmq}, we consider the isospin decomposition
\begin{equation}
\amu^{{\HVP}} = \amu^{\HVP}(I=1) + \amu^{\HVP}(I=0)\,,
\end{equation}
where
\begin{equation}
 \amu^{\HVP}(I=1) = \frac{9}{10}\, \amu^{\HVP}(\ell)
\end{equation}
is the isovector contribution, directly related to the light-quark one, and
\begin{equation}
\amu^{\HVP}(I=0) = \frac{1}{10}\, \amu^{\HVP}(\ell) + \amu^{\HVP}(s) + \amu^{\HVP}(c) + \amu^{\HVP}(\mathrm{disc})
\label{eq:I0_def}
\end{equation}
is the isoscalar contribution, which includes flavour-diagonal and mixed disconnected terms.

\begin{table}[t]
\centering
\footnotesize
\begin{tabular}{lrccccc}
\toprule
ID     & $L~[\fm]$ & $a^{\iso}~[\fm]$   & $am_{\ell}^{\iso}$ & $M_{\pi^0}^{\mathsf{sim}}~[\mev]$ & $Z_V$           & $Z_A$          \\
\midrule 
B64    & $5.09$    & ~ $0.07948(11)$    &  0.0006669(28)     &   109.6             &  $0.706354(54)$ &  $0.74296(19)$ \\
B96    & $7.63$    & ~ $0.07948(11)$    &  0.0006669(28)     &   109.6             &  $0.706406(52)$ &  $0.74261(19)$ \\[2pt]
C80    & $5.45$    & ~ $0.06819(14)$    &  0.0005864(34)     &   116.8             &  $0.725440(33)$ &  $0.75814(13)$ \\[2pt]
D96    & $5.46$    & ~ $0.05685(09)$    &  0.0004934(24)     &   122.7             &  $0.744132(31)$ &  $0.77367(10)$ \\[2pt]
E112   & $5.48$    & ~ $0.04892(11)$    &  0.0004306(23)     &   126.0             &  $0.758238(18)$ &  $0.78548(09)$ \\
\bottomrule
\end{tabular}
\caption{Gauge ensembles generated by the ETM Collaboration used in this work. The lattice spacings and bare light-quark masses are fine-tuned \cite{ExtendedTwistedMass:2024nyi} to match the target isoQCD definition corresponding to the Edinburgh/FLAG consensus \cite{FlavourLatticeAveragingGroupFLAG:2024oxs}.}
\label{tab:iso_EDI_FLAG}
\end{table}

Before discussing the analysis of the isovector and isoscalar contributions, we describe the blinding strategy, which has become a common practice in lattice determinations of $\amu^{\HVP}$ to avoid unconscious bias in the determination of uncertainties. Independent analyses have been performed by three groups on vector--vector two-point functions to which an additive blinding has been applied,
$C^{\mathrm{reg}}(t) \rightarrow C^{\mathrm{reg}}(t) + C^{\mathsf{blind}}_{\mathrm{lat}}(t)$.
The additive blinding correlator is constructed to be identical, after proper current renormalisation, for $\tm$ and $\OS$ regularisations at all lattice spacings and volumes. It is first defined in the continuum as
\begin{align}
C^{\mathsf{blind}}(t) = \sum_{k=1}^{K} a_{k}\, e^{-m_{k}T/2}\,
\cosh\!\left[m_{k}\!\left(\frac{T}{2} - t\right)\right],
\end{align}
where $K \ge 12$ is the number of states included and the coefficients $a_k$ are chosen positive to ensure a positive-definite blinded correlator. The corresponding lattice correlator is
\begin{equation}
C^{\mathsf{blind}}_{\mathsf{lat}}(t) = \frac{a^3}{Z_\mathsf{reg}^{2}}\, C^{\mathsf{blind}}(t)\,,
\end{equation}
with $a$ the lattice spacing and $Z_\mathrm{reg}$ the appropriate vector-current renormalisation constant. This blinding procedure preserves both finite-volume and cutoff effects, enabling the subtraction of perturbative cut-off effects using free theory calculations. Furthermore, after finalising the analysis, one can unblind just by subtracting the quantity
$\Delta_{\mathsf{blind}} \amu^{\HVP} = 2\alpha^2 \int_{0}^{+\infty} \dd{t} t^2\, K(m_{\mu}t)\, C^{\mathsf{blind}}(t)$ from the continuum limit result.

In the following, we present preliminary results obtained by one of the analysis groups.

\section{\texorpdfstring{$I=1$}{I=1} contribution to \texorpdfstring{$\amu$}{amu}}

A precise determination of $\amu^{\HVP}(I=1)$ requires reliable control of the vector--vector correlator at large Euclidean times. Although the signal-to-noise ratio decreases exponentially, the integration kernel enhances the long-distance contribution. It is therefore essential to improve the correlator in the long-distance region and to control the impact of spurious statistical fluctuations.

\begin{wrapfigure}{l}{0.47\textwidth}
\centering
\vspace{-1.3em}
\includegraphics[width=0.47\textwidth]{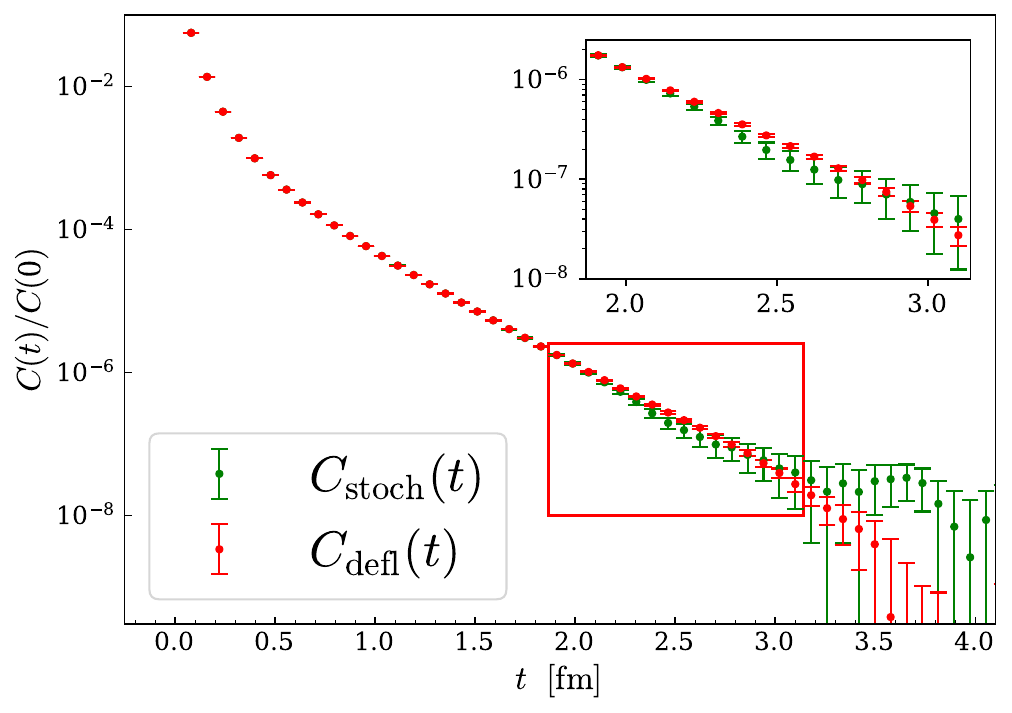}
\caption{On the B64 ensemble, comparison between the standard vector--vector correlator computed fully stochastically (green points) and the LMA-improved one (red points).}
\label{fig:gain}
\end{wrapfigure}
To enhance the signal at large times, we employ Low-Mode Averaging~\cite{Neff:2001zr,Giusti:2004yp,DeGrand:2004qw} (LMA). We compute a set of low-lying eigenpairs $(\ket{v}, \lambda)$ of the Dirac operator and construct a projector onto the infrared part of its spectrum, $P_{\mathsf{IR}}$. The improved correlator is
\begin{equation}
    C^{\mathsf{LMA}}(t) = C_{\eta}(t) - C_{\eta}^{\mathsf{IR}}(t) + C^{\mathsf{IR}}(t)\,,
\end{equation}
where $C_{\eta}(t)$ is obtained from standard stochastic sources, $C_{\eta}^{\mathsf{IR}}(t)$ from stochastic sources projected with $P_{\mathsf{IR}}$, and $C^{\mathsf{IR}}(t)$ from the exact low-mode contribution. As illustrated in \cref{fig:gain} and also shown in Ref.~\cite{ExtendedTwistedMass:2025tpc}, this procedure improves the signal-to-noise ratio of $C(t)$ in the time window dominating the long-distance contribution to $\amu^{\HVP}(I=1)$ by a factor of $\sim 3.5$--$4$.

To further improve the determination of $\amu^{\HVP}(I=1)$, we adopt the bounding procedure proposed in Ref.~\cite{RBC:2018dos}. In this approach, the tail of the correlator for $t \ge t_c$ is replaced by a single-exponential behaviour. The correlator is bounded from above by the lowest-energy state contributing to the channel ($E_0$) and from below by the effective mass ($E_{\mathrm{eff}}$), namely
\begin{equation}
    C(t_c)e^{-E_{\mathrm{eff}}(t_c)\,(t-t_c)} \leq C(t) \leq C(t_c)e^{-E_0\,(t-t_c)}\,,
    \label{Eq:corr_bound}
\end{equation}
In practice, the effective mass entering the lower bound can be evaluated at a fixed $t^\star$ smaller than the first $t_c$ considered.

\begin{figure}[htb!]
\centering
\includegraphics[width=0.49\textwidth]{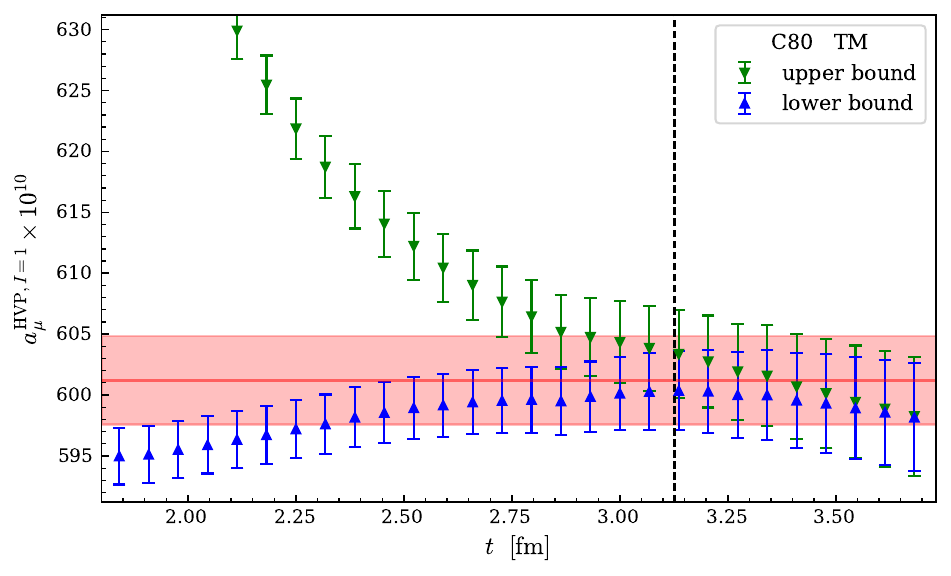}
\includegraphics[width=0.49\textwidth]{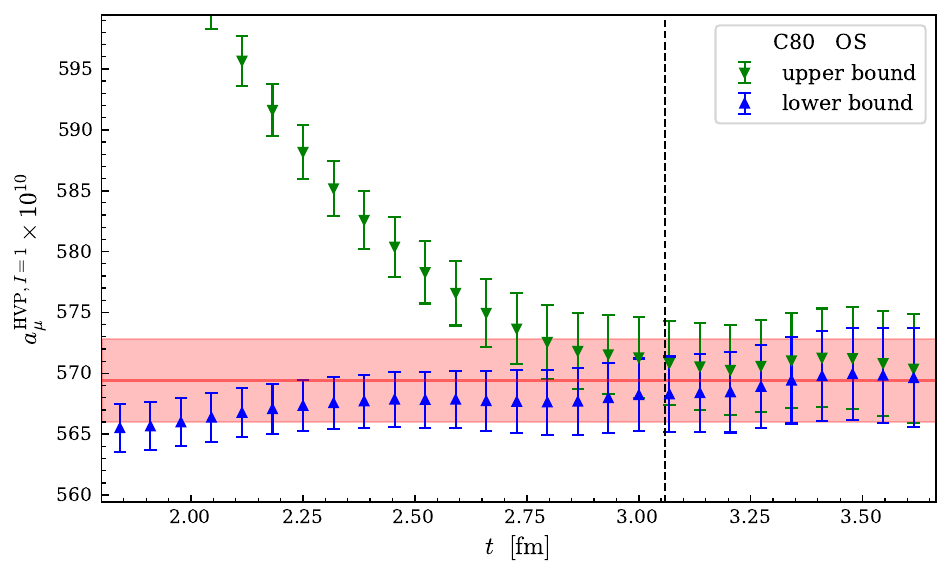}
\caption{Bounding procedure applied to the $I=1$ contribution to $\amu^{\HVP}$ on the C80 ensemble for the $\tm$ (left) and $\OS$ (right) regularisations. For each $t_c$, green points correspond to the upper bound and blue points to the lower bound. The red horizontal line and band show the result obtained by averaging the data points between $t_c^{\mathsf{opt}}$ (vertical dashed line) and $t_c^{\mathsf{opt}} + 0.25~\fm$.}
\label{fig:I1_bound}
\end{figure}
In the isovector channel, for each regularisation and ensemble we use $t^\star \simeq 1.8~\fm$ for the effective mass entering the lower bound. Charge conjugation ($\mathcal{C}$), parity ($\mathcal{P}$), and isospin symmetry imply that the lowest-energy contribution arises from two-pion states with relative momentum, dominated by the $\rho$ resonance. The combination of LMA and the bounding strategy stabilises the long-time behaviour of the correlator and enables a reliable determination of $\amu^{\HVP}(I=1)$ with controlled statistical uncertainties.

\begin{wrapfigure}{r}{0.55\textwidth}
\centering
\vspace{-1.3em}
\includegraphics[width=0.55\textwidth]{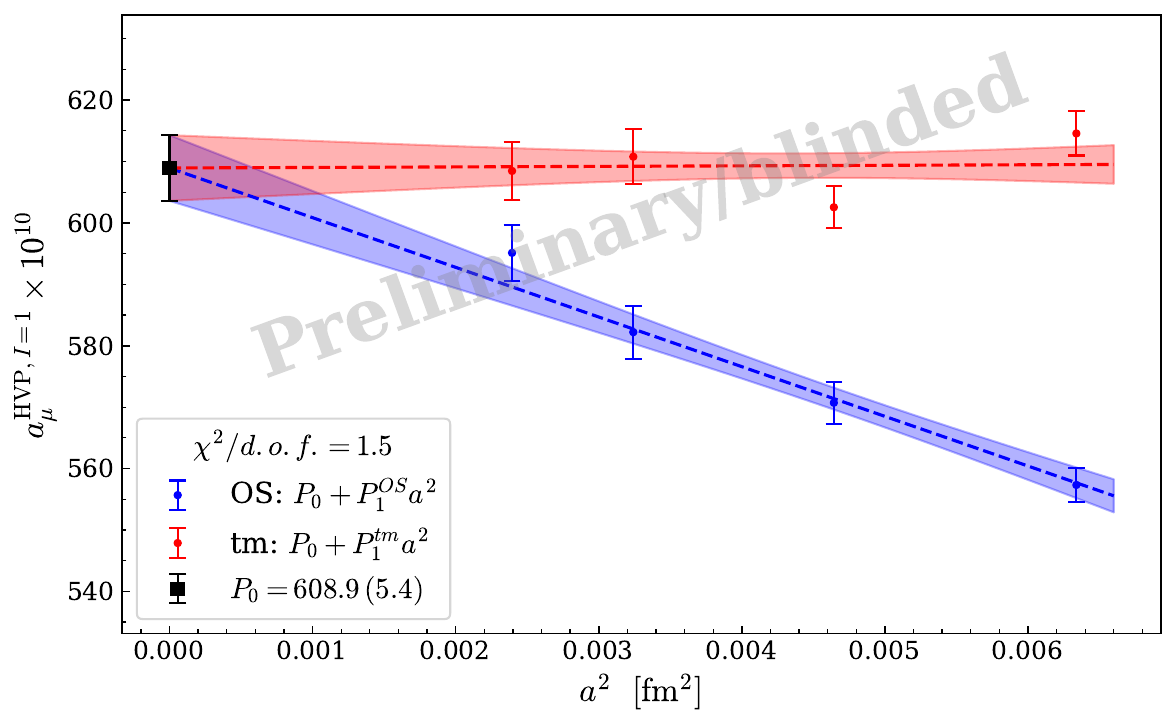}
\caption{Continuum linear--linear combined extrapolation in $a^2$ of $\amu^{\HVP}(I=1)$. Red points correspond to the $\tm$ regularisation, blue points to the $\OS$ one.}
\label{fig:I1_cont}
\end{wrapfigure}
Using the bounded correlator in \cref{eq:amu_def}, one obtains for each $t_c$ a bounded estimate of $\amu^{\HVP}$. For each lattice spacing and regularisation, our final value is obtained by averaging over the time window $\left[t_c^{\mathsf{opt}},\, t_c^{\mathsf{opt}} + 0.25~\fm\right]$. The optimal $t_c$ is defined by the condition that the central values of the upper and lower bounds differ by less than the standard deviation of both bounds, i.e. when the difference of the mean values of the two bounds is less than the minimum of the errors. In \cref{fig:I1_bound}, we show the bounding procedure on the C80 ensemble ($a\simeq0.068~\fm$, $L\simeq 5.45~\fm$) for both $\tm$ and $\OS$ regularisations. This procedure yields absolute errors of $\sim (3.5$--$4)\times 10^{-10}$ across all ensembles. In \cref{fig:I1_cont}, we display the scaling of our blinded data towards the continuum. While the $\tm$ data exhibit a rather flat behaviour, compatible with a constant within errors, we also show a tentative combined fit linear in $a^2$ to both the $\tm$ (red points) and $\OS$ (blue points) data.

\begin{figure}[htb!]
\centering
\includegraphics[width=0.49\textwidth]{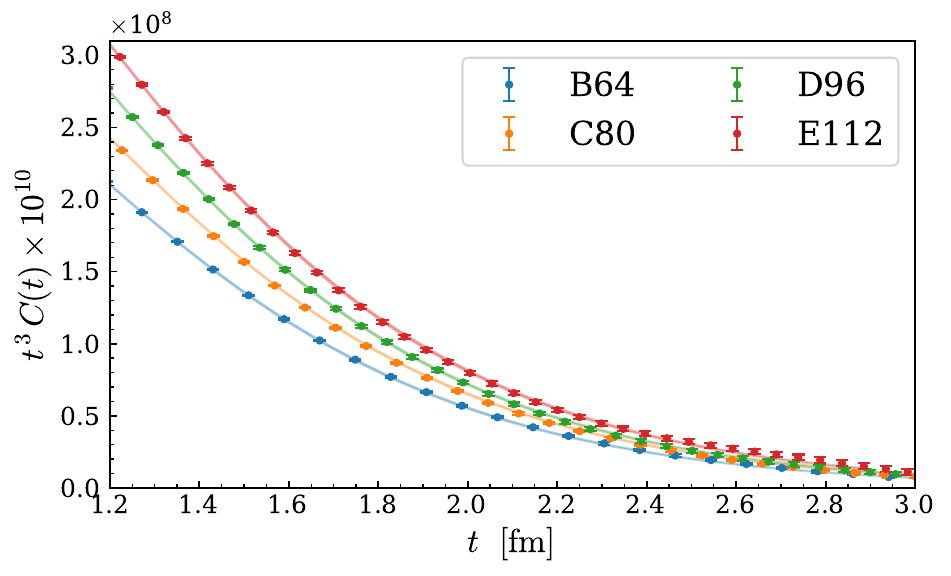}
\includegraphics[width=0.49\textwidth]{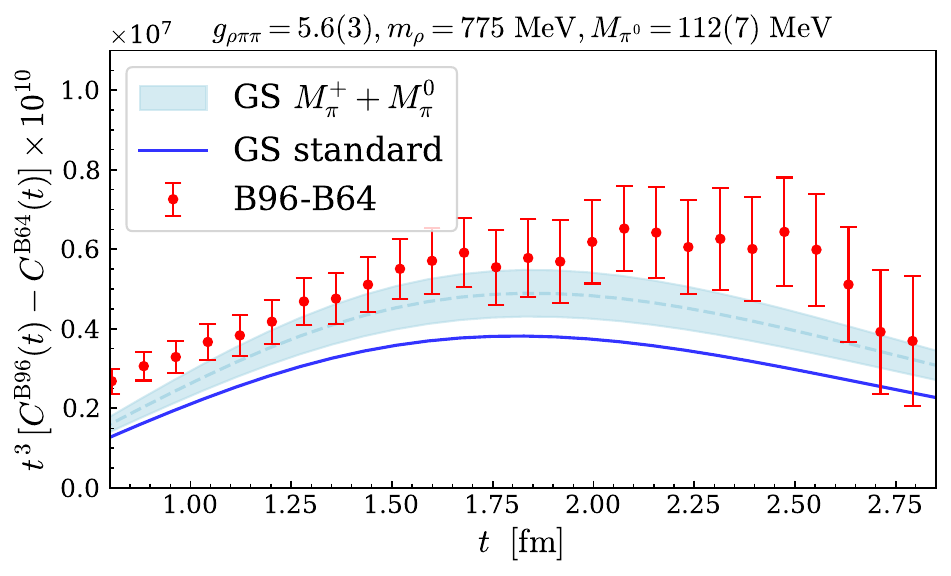}
\caption{Difference between the B96 and B64 correlators (red points) compared with the GS model: standard (dark blue solid line) and modified (light blue band).}
\label{fig:I1_FVE}
\end{figure}

Finally, as indicated by other collaborations~\cite{Boccaletti:2024guq, Djukanovic:2024cmq, RBC:2024fic}, finite-volume effects (FVE) in the $I=1$ contribution are expected to be non-negligible and must be carefully quantified. To this end, we employ the Gounaris--Sakurai (GS) model~\cite{Gounaris:1968mw, Luscher:1985dn, Luscher:1986pf, Luscher:1990ux, Luscher:1991cf, Meyer:2011um, Francis:2013fzp}. To validate this approach, we analyse data from two spatial volumes, $L\simeq 5.09~\fm$ (B64) and $L\simeq 7.63~\fm$ (B96), at our coarsest lattice spacing. Owing to the intrinsic isospin-breaking effects of the twisted-mass action, the validation is performed using a modified version of the GS model that incorporates these effects, which vanish in the continuum limit.
Specifically, previously determined $\pi^{\pm}$ and $\pi^0$ masses are used to compute the energy levels (see \cref{tab:iso_EDI_FLAG}), while the coupling $g_{\rho\pi\pi}$, the $\rho$ mass $M_{\rho}$, and the dual model parameters are extracted from a combined fit to vector--vector correlators across four $\beta$ values. Figure~\cref{fig:I1_FVE} illustrates the results: the left panel shows the outcome of the combined fit, while the right panel compares the standard GS model (dark blue solid line) and the modified GS model (light blue band) with the correlator differences (red points) for B64 and B96. The modified GS model reproduces the data at the $\sim 10$--$15\%$ level, thereby validating the continuum GS model at this level of precision.

\section{\texorpdfstring{$I=0$}{I=0} contribution to \texorpdfstring{$\amu$}{amu}}

While the strange and charm contributions to $\amu^{\HVP}(I=0)$ have been determined with high precision in Ref.~\cite{ExtendedTwistedMass:2024nyi}, the dominant contribution to its uncertainty arises from quark-disconnected diagrams. Since the charmed disconnected contributions (both diagonal and mixed) are expected to be negligible, we focus on the \textit{light--strange} sector. By exploiting the space--time integrated light--strange axial Ward identity (WI)\footnote{
Using the regularisation in which $r_{\ell}=-r_s$, the right-hand side of \cref{Eq:disco_WI} is properly renormalised by the renormalisation constant $Z_V^0$. Moreover, Ref.~\cite{ExtendedTwistedMass:2022jpw} shows that $Z_V = Z_V^0$.
}
\begin{align}
\sum_{f=u,d,s} q_{f} B^{\mu}_{f}(z)
= Z_V (m_{\ell} + m_{s}) a^4 \sum_{x} \expval{P_{\ell s}(x) A^{\mu}_{\ell s}(z)}\,,
\quad
B^{\mu}_{f}(z) = \expval{\Tr{\Hat{V}_f^{\mu}(z)}} + \order{a}\,,
\label{Eq:disco_WI}
\end{align}
where $\Hat{V}_f^{\mu}(z)$ is the properly renormalised flavour-$f$ current, we can compute the Wick contractions of the disconnected $u$-, $d$-, and $s$-quark contributions in isoQCD.

This method generalises the one-end trick commonly used with twisted-mass fermions. Moreover, as argued in Ref.~\cite{Giusti:2019kff} based on spectral-decomposition arguments, the variance of differences is expected to be significantly smaller than for standard estimators of differences of single-propagator traces. To further improve the signal of quark-disconnected correlators, we also employ the frequency-splitting technique proposed in Ref.~\cite{Giusti:2019kff},
\begin{multline}
B^{\mu}(\mu_{\ell}) - B^{\mu}(\mu_{s})
= \left[ B^{\mu}(\mu_{\ell}) - B^{\mu}(\mu_{1}) \right]
+ \left[ B^{\mu}(\mu_{1}) - B^{\mu}(\mu_{2}) \right] \\
+ \dots +
\left[ B^{\mu}(\mu_{N}) - B^{\mu}(\mu_{s}) \right] .
\end{multline}
By decomposing the full difference into a sum of differences that probe distinct frequency regions, one achieves a substantial variance reduction. In practice, we find that a four-level splitting ($N=3$) provides an optimal balance between computational cost and statistical precision.

We apply the same bounding strategy used in the isovector analysis to the $\amu^{\HVP}(I=0)$ contribution (see \cref{eq:I0_def}). At this stage of the analysis, we consider only the non-charm contributions. Indeed, while we already computed the quark-connected contribution with percent precision~\cite{ExtendedTwistedMass:2024nyi}, the quark-disconnected charm contributions (diagonal and mixed) are expected to be negligible within our uncertainties.
Symmetry arguments based on $\mathcal{C}$, $\mathcal{P}$, and isospin imply that the leading intermediate states are three-pion configurations, dominated by the $\omega$ resonance. Accordingly, we employ positivity of the correlator for the lower bound and the lowest three-pion energy,
\begin{equation}
E_0 = 2\sqrt{m_{\pi^{\pm}}^2 + \vb*{p}^{\,2}} + m_{\pi^0},
\end{equation}
for the upper bound. Here, it is important to stress that this three-pion state can occur only with parity-breaking regularisation such as the twisted-mass one. Indeed, in the continuum, the neutral pion should also have momentum such that the total momentum of the system is zero. 

For each lattice spacing, the final result is obtained by averaging the bounded estimates over the interval $\left[t_c^{\mathsf{opt}},\, t_c^{\mathsf{opt}} + 0.25~\fm\right]$, where $t_c^{\mathsf{opt}}$ is determined using the same stability criterion adopted in the $I=1$ channel. The resulting data exhibit a weak dependence on the lattice spacing, indicating small cutoff effects. To avoid underestimating systematic uncertainties, we perform both a linear fit in $a^2$ using all ensembles and a constant fit using the three finest lattice spacings. The final estimate is obtained from a Bayesian AIC combination of these fits. Finite-volume effects are found to be negligible within our precision, as indicated by the agreement between the B64 and B96 ensembles. Given the dominance of three-pion intermediate states, such suppression is expected and consistent with observations from other collaborations~\cite{Djukanovic:2024cmq}.
\begin{figure}[htb]
    \centering
    \includegraphics[width=0.49\linewidth]{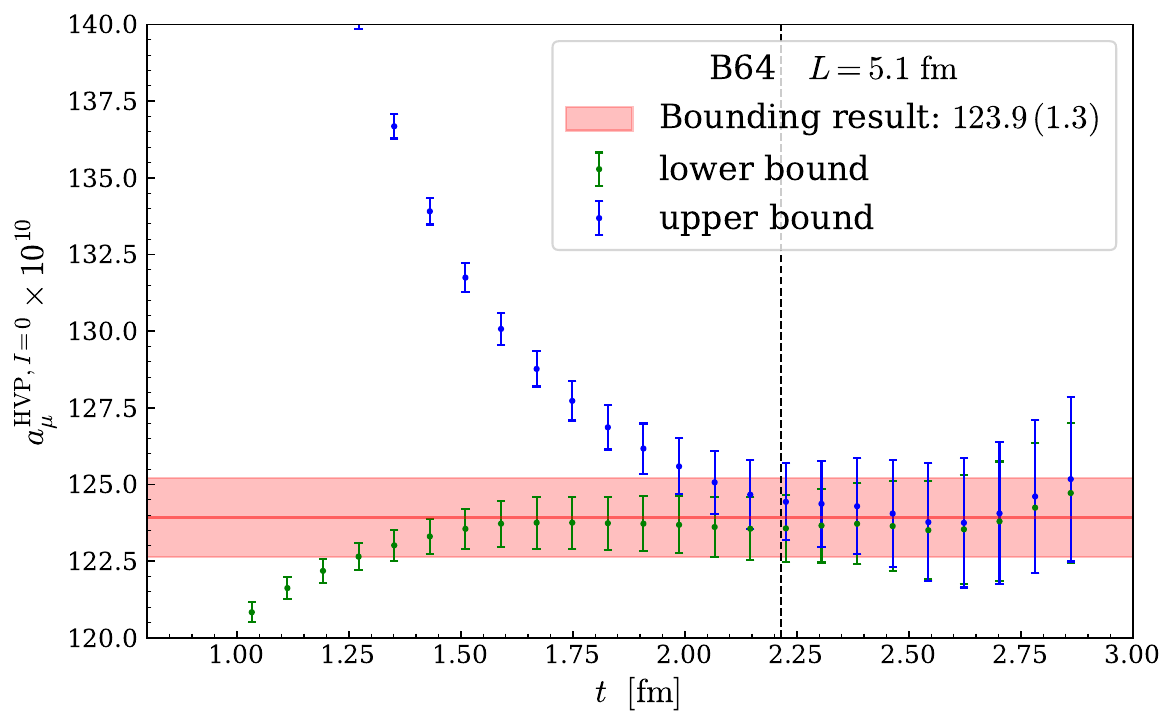}
    \includegraphics[width=0.49\linewidth]{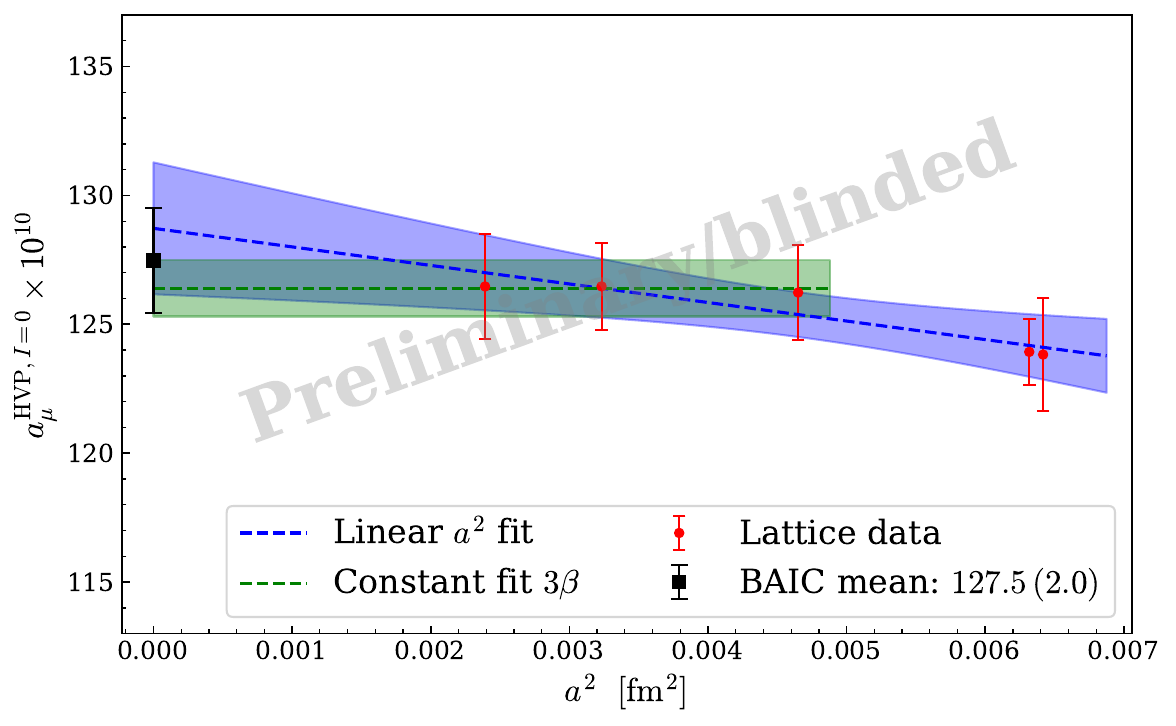}
    \caption{\textit{Left}: Bounding procedure for $\amu^{\HVP}(I=0)$ on the B64 ensemble. For each $t_c$, green points correspond to the upper bound and blue points to the lower bound. The red horizontal line and band show the result obtained by averaging the data points between $t_c^{\mathsf{opt}}$ (vertical dashed line) and $t_c^{\mathsf{opt}} + 0.25~\fm$. \textit{Right}: Preliminary continuum extrapolation. Red points correspond to the data obtained on the ensembles of \cref{tab:iso_EDI_FLAG}; the green band represents a constant fit to the three finest ensembles, while the blue band shows a linear fit to all ensembles. The black point denotes the result obtained from a Bayesian AIC combination of these fits.}
    \label{fig:I0_contr}
\end{figure}

In \cref{fig:I0_contr}, we display the bounding procedure on the B64 ensemble ($a\simeq 0.079~\fm$, $L\simeq 5.10~\fm$) together with a preliminary continuum extrapolation. The data exhibit a flat behaviour, indicating small discretisation effects. To avoid underestimating systematic uncertainties, we perform both a linear fit using all lattice spacings and a constant fit using the three finest ensembles, and combine them via a Bayesian AIC procedure~\cite{Neil:2022joj}. Finite-volume effects are negligible within our precision, as indicated by the two rightmost points in the right panel of \cref{fig:I0_contr}, corresponding to the B64 and B96 ensembles. Given the dominance of three-pion intermediate states, such effects are expected to be suppressed, in agreement with findings of other collaborations.

\section*{Acknowlogments}
{\small
We thank all members of ETMC for the most enjoyable collaboration. 
A.E., S.B. and F.P. acknowledge support from EXCELLENCE/0524/0017 (MuonHVP) and EXCELLENCE/0524/0459 (IMAGE-N), co-financed by the European Regional Development Fund and the Republic of Cyprus through the Research and Innovation Foundation within the framework of the Cohesion Policy Programme “THALIA 2021-2027”.
A.E., R.F., G.G., L.M. and F.M. are supported by the Italian Ministry of University and Research (MUR) under the grant PNRR-M4C2-I1.1-PRIN 2022-PE2 Non-perturbative aspects of fundamental interactions, in the Standard Model and beyond F53D23001480006 funded by E.U.- NextGenerationEU. 
M.G. is supported by the Deutsche Forschungsgemeinschaft (DFG, German Research Foundation) as part of the CRC 1639 NuMeriQS – project no. 511713970.
This work was supported by the Swiss National Science Foundation (SNSF) through the grants \href{https://data.snf.ch/grants/grant/208222}{208222} and \href{https://data.snf.ch/grants/grant/10003675}{10003675}.
We gratefully acknowledge CINECA and EuroHPC Joint Undertaking for granting access to the Leonardo Supercomputer. Computing time on Leonardo Booster was allocated through the Extreme Scale Access Call (grant EHPC-EXT-2024E01-027), and additional GPU resources were provided under the INFN-LQCD123 initiative.
We acknowledge the Swiss National Supercomputing Centre (CSCS) access to Alps through the Chronos programme under project ID CH15 as well as CSCS and the EuroHPC Joint Undertaking for awarding  access to the LUMI supercomputer, owned by the EuroHPC Joint Undertaking, hosted by CSC (Finland) and the LUMI consortium through the Chronos programme under project ID CH17-CYP. The authors acknowledge the Texas Advanced Computing Center (TACC) at The University of Texas at Austin for providing HPC resources (Project ID PHY21001).
The authors gratefully acknowledge the Gauss Centre for Supercomputing e.V. (\href{https://www.gauss-centre.eu/}{www.gauss-centre.eu}) for funding this project by providing computing time on the GCS  Supercomputer JUWELS \cite{JUWELS} at J\"ulich Supercomputing Centre (JSC).
}

\bibliographystyle{JHEP}
\nocite{*}
\bibliography{biblio.bib}

\end{document}